# Leveraging Global Binary Masks for Structure Segmentation in Medical Images

Mahdieh Kazemimoghadam[1], Zi Yang[1], Mingli Chen[1], Lin Ma[1], Weiguo Lu[1*] and Xuejun Gu[1,2*]

[1]Department of Radiation Oncology, the University of Texas Southwestern Medical Center, Dallas TX, 75390 USA

[2]Department of Radiation Oncology, Stanford University, Stanford, CA 94305

*E-mails: weiguo.lu@utsouthwestern.edu, xuejun.gu@utsouthwestern.edu, xuejungu@stanford.edu

**Abstract**

Deep learning (DL) models for medical image segmentation are highly influenced by intensity variations of input images and lack generalization due to primarily utilizing pixels' intensity information for inference. Acquiring sufficient training data is another challenge limiting models' applications. Here, we proposed to leverage the consistency of organs' anatomical position and shape information in medical images. We introduced a framework leveraging recurring anatomical patterns through global binary masks for organ segmentation. Two scenarios were studied: (1) Global binary masks were the only input for the U-Net based model, forcing exclusively encoding organs' position and shape information for rough segmentation or localization. (2) Global binary masks were incorporated as an additional channel providing position/shape clues to mitigate training data scarcity. Two datasets of the brain and heart CT images with their ground-truth were split into (26:10:10) and (12:3:5) for training, validation, and test respectively. The two scenarios were evaluated using full training split as well as reduced subsets of training data. In scenario (1), training exclusively on global binary masks led to Dice scores of 0.77±0.06 and 0.85±0.04 for the brain and heart structures respectively. Average Euclidian distance of 3.12±1.43mm and 2.5±0.93mm were obtained relative to the center of mass of the ground truth for the brain and heart structures respectively. The outcomes indicated encoding a surprising degree of position and shape information through global binary masks. In scenario (2), incorporating global binary masks led to significantly higher accuracy relative to the model trained on only CT images in small subsets of training data; the performance improved by 4.3-125.3% and 1.3-48.1% for 1-8 training cases of the brain and heart datasets respectively. The findings imply the advantages of utilizing global binary masks for building models that are robust to image intensity variations as well as an effective approach to boost performance when access to labeled training data is highly limited.

## 1. Introduction

In recent years, deep learning (DL) approaches have been successfully utilized for medical image segmentation. Current deep learning methods mainly rely on quantitative visual appearance of input images which is typically associated with individual pixel- or voxel-wise intensities. The models perform intensity-based segmentation where pixel intensity information is primarily utilized for inference. Despite their success, such intensity-based segmentation is highly influenced by variations in input images. The image intensity distribution shift (or domain shift) between the training and the testing data usually results in severe performance degeneration during the deployment of the trained models into practical use (Yasaka and Abe 2018). Such intensity distribution shift typically come from different acquisition parameters, various imaging methods or diverse modalities. For instance, in clinical practice, images from different modalities are frequently used to capture anatomical structures. Computed tomography (CT) and magnetic resonance (MR) are examples of widely used modalities providing clear images of different organs such as heart, brain, and kidneys with high spatial resolution.

Imaging modalities utilize different physical principles, leading to distinct visual appearance and organs' contrast (Shung, K. Kirk, Michael B. Smith 2012). Hence, for instance, for algorithms trained to segment the brainstem in MRI, it is quite impossible to segment the same structure in CT due to the fact that the learning step of the algorithm is performed on a different distribution than the targeted application. This introduces one limitation of current DL models which is the lack of generalization across image variations (Zhang *et al* 2020). Thus, one of the challenges in medical images segmentation is the need to find common shared representations across images with different visual appearance. We postulate that the consistent anatomical shape and position information naturally embedded in medical images can provide common shared representations. Leveraging such recurring anatomical patterns is a promising direction to build generalizable models for organ localization.

In recent years, studies have reported on utilizing spatial positioning and shape information by CNN networks for decision making and to assist with image segmentation. The ability of CNNs to implicitly encode target position and shape has been demonstrated by providing examples of trained networks on standard tasks. For instance, direct access to image position through an additional channel containing embedded image coordinate was shown to improve the segmentation of medical images and the robustness of the model in different circumstances (Murase 2020). Incorporating such positional information was shown to increase robustness of shallow CNN networks and when training data was limited. Other approaches such as designing a loss function with specific penalty terms for desired shape priors were also introduced to encode topological information (BenTaieb and Hamarneh 2016). In (Larrazabal *et al* 2020), the advantages of leveraging shape and topological priors for improving the quality of noisy segmentation masks of chest X-ray images was reported. The authors introduced a novel approach independent of image intensity by post-processing the predicted mask of an initial segmentation model. (Oktay *et al* 2018) presented a generic approach to incorporate global shape and anatomy information into deep learning models. Their approach learned representations of objects' shape and location and enforced the model to predict segmentation masks that followed the learnt shape priors.

On the other hand, despite emerging numerous state-of-the-art deep learning models for medical image segmentation, the models' applicability in clinical settings has been limited due to the heavy reliance on large amounts of training data, especially annotated images. In order to minimize model overfitting and to achieve accurate outcomes, access to large training datasets is necessary. However, it is often challenging to acquire sufficient manually annotated training data as annotation of medical images are costly and highly time and labor intensive. Approaches such as data augmentation (Chaitanya *et al* 2021a), transfer learning (Chaitanya *et al* 2021b), semi-supervised (You *et al* 2022) and self-supervised learning (Haghighi *et al* 2021) have been reported to handle training annotation scarcity. However, the proposed models still require tens to hundreds of images/ground truth masks to train. Hence, preparing training data for DL models is still a substantial burden especially where the regions of interests are complex as annotating a single image could take hours for complex target regions with a large number of structures (Brunenberg *et al* 2020, Zhuang and Shen 2016). We further postulate that the global binary mask implicitly encoding the consistent anatomical shape and position information can be incorporated to build accurate models for medical image segmentation. This particularly leads to performance improvement where access to training data is highly limited.

Our contribution in this paper is two folds: 1) we leverage the consistency of organs' anatomical position and shape across patients to build generalizable organ localization and rough segmentation models. Our proposed approach utilizes global binary masks to force the model exclusively learn anatomical position and shape information rather than intensity for organ segmentation. We hypothesize that the global binary masks can provide the means for the deep learning network to encode organs' anatomical position and shape for medical image segmentation. Our approach is independent of image modality and intensity information as it employs only global binary masks for training. 2) The global binary masks are incorporated as an additional channel providing position/shape clues to improve segmentation outcomes where access to training data is highly limited. We hypothesize that incorporating prior knowledge in the form of global binary masks can compensate for scarcity of training data. The effectiveness of the proposed approaches using varying subsets of training data is then demonstrated.

## 2. Methods and Materials
### 2.1. Leveraging recurring anatomical patterns for medical image segmentation
We introduced a framework taking the advantage of consistent anatomical patterns embedded in medical images to encode and provide organs' position and shape information for medical image segmentation. We studied two scenarios: (1) Global binary masks served as the only input for the segmentation model (Figure 1); (2) Global binary masks were incorporated as an additional channel (Figure 2). In the first scenario, our goal was to build a model independent of image intensity and modality. For this purpose, the model was forced to exclusively learn structures/organs shape and position rather than intensity, as the binary masks were the only inputs for the network. In the second scenario, our proposed approach incorporated anatomical prior information through global binary masks into the deep learning model for image segmentation. The model utilized pre-generated global binary masks as an additional channel to provide position/shape clue for the segmentation network. We trained the model on CT images along with the corresponding global binary masks. The outcomes of the proposed two scenarios were subsequently compared with the model trained on CT images only.

We employed U-Net (Ronneberger *et al* 2015) as the backbone of our proposed method consisting of an up-sampling path and a down-sampling path for extracting contextual and spatial information respectively. The down-sampling or contracting path was comprised of four convolutional blocks each followed by a 3D max-pooling with the kernel size of 2. The first block had 32 channels. The number of channels was doubled after every max-pooling layer. The up-sampling or expanding path was also composed of four convolutional blocks to up-sample feature maps. The number of channels was halved in every step of the expanding path. Skip connections were applied to concatenate feature maps of up-sampling and down-sampling paths. All the convolutional blocks included two consecutive convolutional layers with kernels of size 3 × 3 × 3, each followed by a batch normalization layer, and a sigmoid activation function. A 3D spatial dropout with the rate of 0.2 was also applied after the second convolution layer of each block.

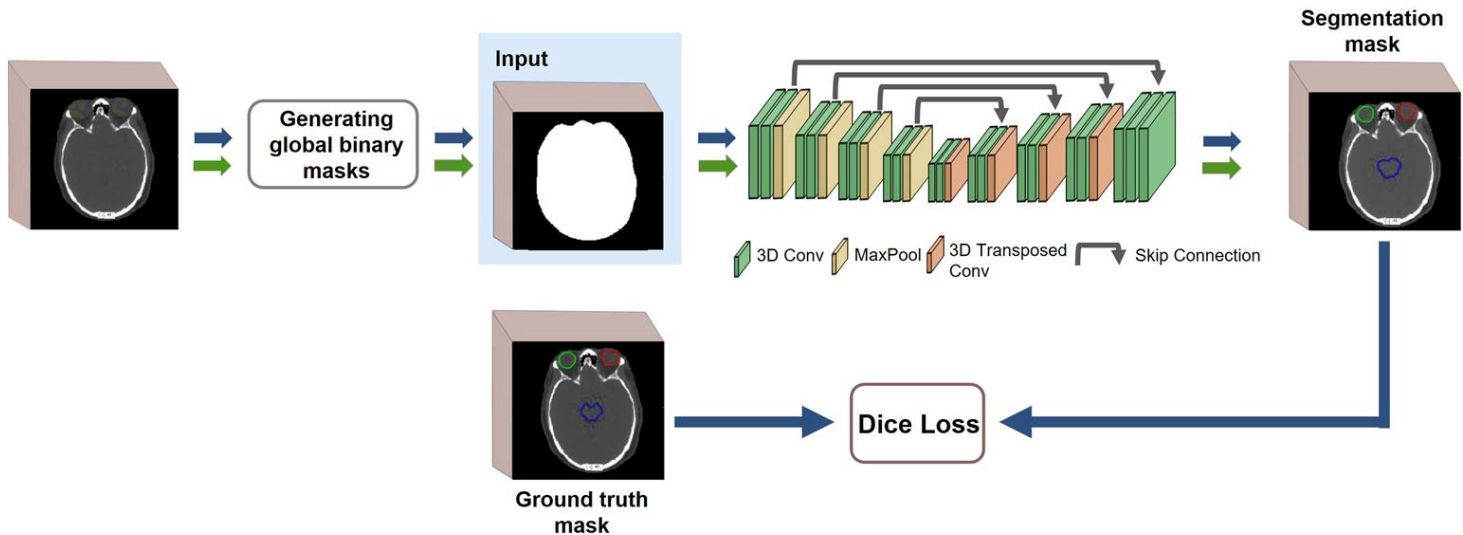

Figure 1. Network structure for the proposed approach in the first scenario, where the model was exclusively trained on global binary masks for structure/organ localization and rough segmentation. Blue and green arrows denote the network training and test steps respectively.

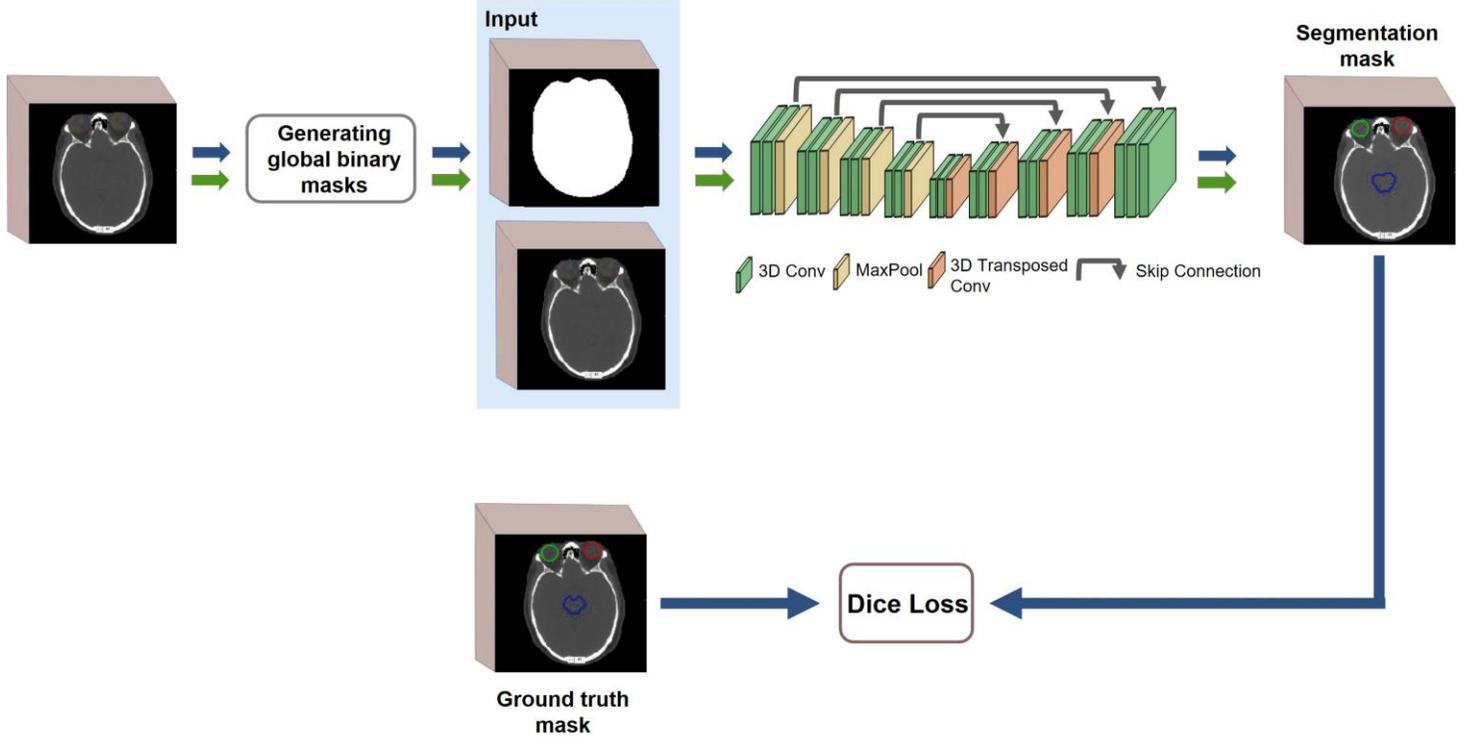

Figure 2. Network structure for the proposed approach in the second scenario, where global binary masks were incorporated as an additional input channel to provide position/shape information for improving segmentation outcomes in small datasets. Blue and green arrows denote the network training and test steps respectively.

### 2.2. Experimental data and model training

Two medical image datasets were used to validate the efficacy of the proposed approaches in our study. For the first dataset, 46 CT images were retrospectively collected from brain cancer patients with the ground truth manual contours for three structures including the brain stem, left eye, and right eye. Manual contours were initially generated by the attending radiation oncologists and later reviewed/refined by neurosurgeons. For this dataset, 26 images were used in total for training, 10 for validation and 10 for test datasets.

The second dataset was provided by the Multi-Modality Whole Heart Segmentation (MMWHS) challenge (Zhuang *et al* 2019). The dataset included cardiac CT scans acquired using two 64-slice CT scanners (Philips Medical Systems, Netherlands) utilizing a standard coronary CT angiography protocol. All images cover the entire heart from the upper abdomen to the aortic arch, with the in-plane resolution of the axial slices of 0.78×0.78 mm, and the average slice thickness of 1.60 mm. We aimed to delineate seven heart substructures including the left ventricle (LV), right ventricle (RV), left atrium (LA), right atrium (RA), myocardium of LV (Myo), ascending aorta (AO) or the whole aorta, and the pulmonary artery (PA). For the MMWHS dataset, 12 images were used in total for training, 3 for validation and 5 for test datasets.

Global binary masks of the head and heart were generated for both datasets. For this purpose, CT images of the head were registered to a template head, then a threshold based segmentation was applied to generate binary masks for the brain dataset. For the heart dataset, a trained U-Net was utilized to generate whole heart binary masks.

The proposed model parameters were optimized by minimizing the Dice loss which compared the predicted masks with manually delineated ground truth labels (1). In equation (1), $Y^P$ and $Y^G$ are the prediction mask and manual segmentation with K voxels, respectively.

$$L_D = 1 - \frac{2\sum_i^k Y_i^P Y_i^G}{\sum_i^k (Y_i^P)^2 + \sum_i^k (Y_i^G)^2} \qquad (1)$$

In order to validate the efficacy of the proposed approaches, the models were trained on the full training split of each dataset as well as on reduced sets of training data. In the first scenario, we aimed to determine the level of position and shape information encoded by the network for varying amounts of training data. In the second scenario, reduced sets of training data were utilized to create room for improvement by global binary masks and verify the effectiveness of the proposed approach for when the training data is insufficient. For the brain dataset, decreasing training size from 26 to 16, 8, 4, 2, and 1 were tested. For the heart dataset, 1,2,4,8, and 12 training cases were examined.

## 2.3. Implementation details

For conducting experiments and implementing our segmentation models, we used Pytorch (1.8) DL library and python (version 3.7) environment on Windows 10, 64x, Intel Xeon processor CPU with 64 GB RAM, and NVIDIA GeForce 2080 Ti GPU with 12 GB memory. To make the CT images resolution consistent among all data, images were resampled to voxel size of 1.5mm×1.5mm×1.5mm and cropped to the image size of 128×128×128. The voxel intensity of CT images was normalized to 0–1 according to HU window [−200 200]. The network was trained with the Adam optimizer and a learning rate of $10^{-4}$ and batch size of 1. The models performed best on validation sets were selected to evaluate the test dataset.

## 3. Results

### 3.1. Utilizing global binary masks to encode anatomical shape/position information

We first examine the effectiveness of the global binary masks for positional and shape encoding. To be specific, we trained a U-Net based model using global binary masks as the only input for both the brain and heart datasets. We then compared the results with those provided by the model trained on CT images only.

### 3.1.1. Qualitative evaluation

Figure 3, rows A and B, present examples of test cases for the model trained exclusively on global binary masks and full training split (i.e., 26 cases) of the brain dataset. The segmentation outcome of the proposed model (third column) was compared with the model trained on CT images (second column) and the ground truth masks (first column). For the CT images displayed in Figure 3, the proposed model provided relatively accurate segmentation of the left and right eyes. The predicted masks for these two structures were comparable to that of trained on CT images and showed high agreement with the ground truth masks. Segmentation of the brain stem appeared to be more challenging for the proposed approach compared to the model trained on CT images, leading to slight over-segmentation and under-segmentation of the brain stem for the two test cases.

Examples of test cases for the model trained exclusively on the full training split of the heart global binary masks (i.e., 12 cases) are presented in Figure 3, rows C and D. Segmentation of the heart substructures using exclusively global binary masks, appeared to be comparable to the model trained on CT images. The outcomes presented high agreement with the ground truth masks. In row C, using the proposed approach, only a slight level of over-segmentation of AA and under-segmentation of RA and RV were observed. The proposed model led to relatively accurate prediction of LA, LV and Myo. In row D, the proposed model provided relatively accurate segmentation of LA, and outperformed the model trained on CT images for segmentation of AA. Only slight levels of over-segmentation were observed for RV, RA, and Myo for the model exclusively trained on global binary masks. Overall, while some degrees of mis-segmentation were observed for both datasets, the predicted masks by the proposed approach matched the results of the model trained on CT images as well as the ground truth masks to a large extent. According to the outcomes, while the proposed approach may not provide detailed segmentation of all the structures, its comparable outcomes relative to the ground truth, confirmed the feasibility of utilizing global binary masks for organ/structure rough segmentation and localization.

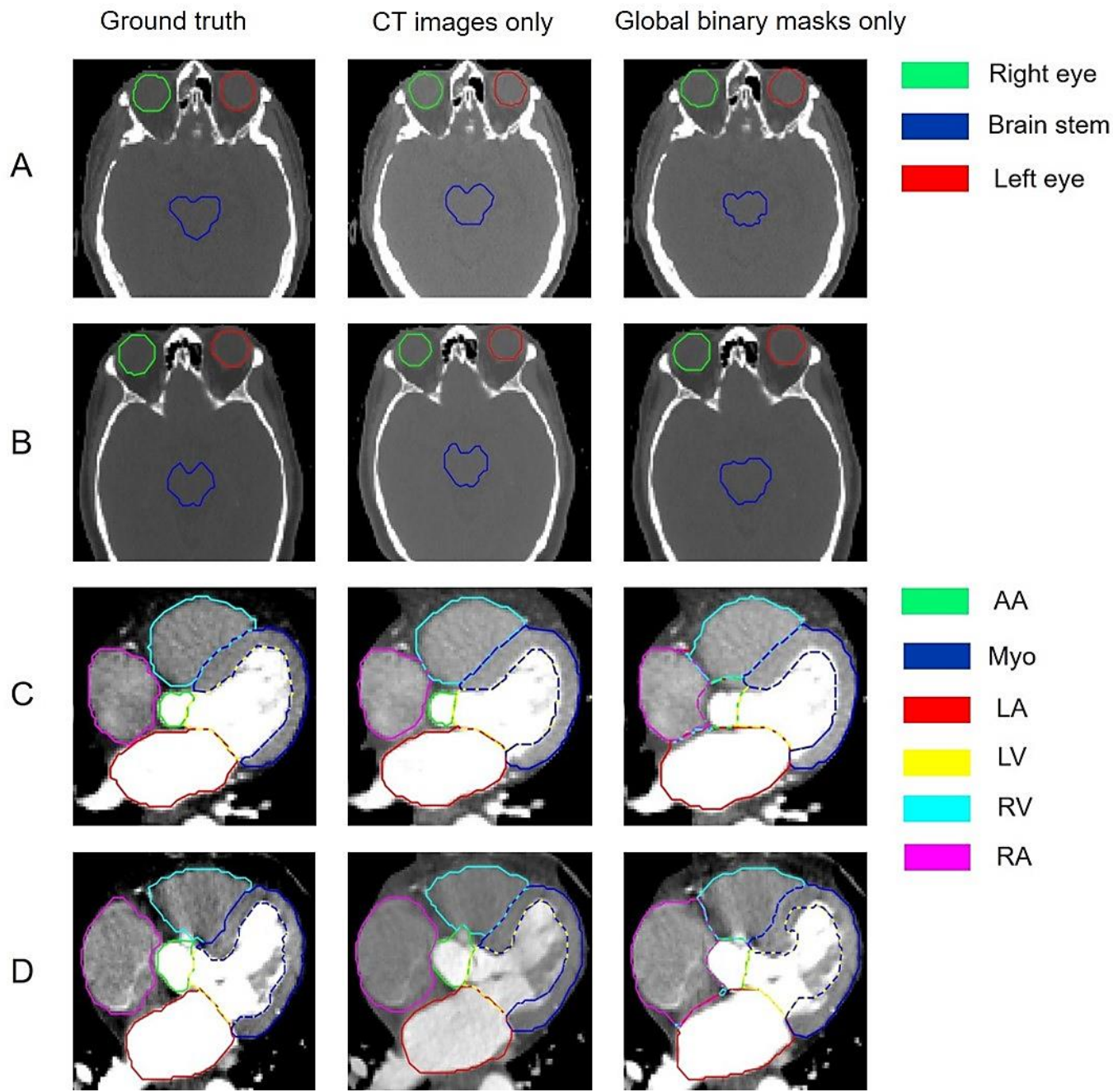

Figure 3. Qualitative comparison of the predicted contours by the model trained with CT images only (second column) and the model trained with global binary masks only (third column), with respect to the ground truth contours (first column). The results for the brain (rows A, B) and heart (rows C, D) datasets are presented. Rows A-D are examples of test cases when the models were trained on the full training split of each dataset (i.e., 26 and 12 cases for the brain and heart datasets respectively).

### 3.1.2. Quantitative results

Utilizing global binary masks as the only input, provided Dice coefficient (DC) of 0.77±0.06 across the brain structures when the model trained on full training set (i.e., 26 cases) (Table 1). For the heart dataset (Table 2), DC of 0.85±0.04 was achieved across structures when the model was exclusively trained on full training set of the heart global binary masks (i.e., 12 cases). This confirmed the fact that a high degree of position and shape information were encoded by the network via global binary masks.

Euclidian distance from the center of mass (COM) of each structure in the predicted mask relative to the ground truth was also calculated as presented in Table 3 and Table 4. Average distance of 3.12±1.43mm was obtained across the brain structures for the model trained on full split of the training data (Table 3). For the heart dataset (Table 4), average distance of 2.5±0.93mm was obtained. The relatively small deviations of predicted masks' COMs with respect to the ground truth confirmed the feasibility of the proposed approach for structure localization purposes in medical images.

Table 1. Dice coefficients for the segmentation masks provided by the model trained on CT images only, trained on global binary masks only (the proposed first scenario), and trained on CT and global binary masks (the proposed second scenario). Dice coefficients were calculated with respect to the ground truth for the subsets of training data for the brain structures. Values in parenthesis are standard deviation.

| Number of training cases | structures | CT images only | Global binary masks only | CT images and global binary masks |
|---|---|---|---|---|
| 1 | Brain stem | 0.4 (0.14) | 0.27 (0.21) | 0.57 (0.09) |
| 1 | Left eye | 0.35 (0.08) | 0.31 (0.25) | 0.66 (0.06) |
| 1 | Right eye | 0 | 0.62 (0.1) | 0.46 (0.16) |
| 2 | Brain stem | 0.6 (0.14) | 0.43 (0.13) | 0.68 (0.07) |
| 2 | Left eye | 0.77 (0.08) | 0.60 (0.12) | 0.74 (0.07) |
| 2 | Right eye | 0 | 0.64 (0.15) | 0.63 (0.1) |
| 4 | Brain stem | 0.57 (0.11) | 0.52 (0.13) | 0.7 (0.05) |
| 4 | Left eye | 0.64 (0.11) | 0.72 (0.1) | 0.75 (0.1) |
| 4 | Right eye | 0.78 (0.06) | 0.78 (0.07) | 0.79 (0.07) |
| 8 | Brain stem | 0.7 (0.07) | 0.62 (0.09) | 0.77 (0.06) |
| 8 | Left eye | 0.78 (0.06) | 0.77 (0.14) | 0.82 (0.04) |
| 8 | Right eye | 0.84 (0.06) | 0.79 (0.09) | 0.83 (0.05) |
| 16 | Brain stem | 0.78 (0.03) | 0.66 (0.09) | 0.8 (0.05) |
| 16 | Left eye | 0.84 (0.05) | 0.79 (0.08) | 0.84 (0.05) |
| 16 | Right eye | 0.84 (0.08) | 0.81 (0.07) | 0.82 (0.06) |
| 26 | Brain stem | 0.82 (0.04) | 0.71 (0.07) | 0.82 (0.05) |
| 26 | Left eye | 0.86 (0.05) | 0.82 (0.07) | 0.87 (0.05) |
| 26 | Right eye | 0.87 (0.06) | 0.79 (0.05) | 0.85 (0.07) |

Table 2. Dice coefficients for the segmentation masks provided by the model trained on only CT images, trained on global binary masks only (the proposed first scenario), and trained on CT and global binary masks (the proposed second scenario). Dice coefficients were calculated with respect to the ground truth for subsets of training data for the heart structures. Values in parenthesis are standard deviation.

| Number of training cases | structures | CT images only | Global binary masks only | CT images and Global binary masks |
|---|---|---|---|---|
| 1 | Myo | 0.59 (0.15) | 0.64 (0.1) | 0.69 (0.11) |
| | LA | 0.6 (0.21) | 0.78 (0.08) | 0.73 (0.18) |
| | LV | 0 (0) | 0.69 (0.05) | 0.66 (0.09) |
| | RA | 0.53 (0.17) | 0.66 (0.13) | 0.73 (0.08) |
| | RV | 0.67 (0.11) | 0.72 (0.11) | 0.77 (0.09) |
| | AA | 0.46 (0.23) | 0.56 (0.21) | 0.6 (0.17) |
| | PA | 0.37 (0.17) | 0.65 (0.1) | 0.59 (0.08) |
| 2 | Myo | 0.58 (0.15) | 0.72 (0.07) | 0.74 (0.11) |
| | LA | 0.81 (0.06) | 0.79 (0.09) | 0.86 (0.07) |
| | LV | 0.83 (0.04) | 0.78 (0.08) | 0.8 (0.1) |
| | RA | 0.66 (0.15) | 0.71 (0.09) | 0.79 (0.05) |
| | RV | 0.71 (0.11) | 0.75 (0.1) | 0.77 (0.11) |
| | AA | 0.67 (0.18) | 0.68 (0.17) | 0.75 (0.18) |
| | PA | 0.63 (0.08) | 0.7 (0.05) | 0.73 (0.05) |
| 4 | Myo | 0.77 (0.08) | 0.78 (0.07) | 0.84 (0.08) |
| | LA | 0.9 (0.06) | 0.9 (0.04) | 0.93 (0.02) |
| | LV | 0.85 (0.05) | 0.8 (0.04) | 0.89 (0.03) |
| | RA | 0.83 (0.06) | 0.86 (0.05) | 0.89 (0.04) |
| | RV | 0.83 (0.05) | 0.89 (0.02) | 0.89 (0.02) |
| | AA | 0.92 (0.04) | 0.8 (0.12) | 0.92 (0.03) |
| | PA | 0.77 (0.13) | 0.79 (0.07) | 0.81 (0.08) |
| 8 | Myo | 0.81 (0.11) | 0.79 (0.11) | 0.85 (0.07) |
| | LA | 0.92 (0.03) | 0.91 (0.03) | 0.94 (0.02) |
| | LV | 0.85 (0.03) | 0.77 (0.06) | 0.91 (0.04) |
| | RA | 0.9 (0.03) | 0.88 (0.04) | 0.91 (0.04) |
| | RV | 0.91 (0.01) | 0.91 (0.02) | 0.9 (0.03) |
| | AA | 0.96 (0.01) | 0.86 (0.07) | 0.96 (0.02) |
| | PA | 0.87 (0.07) | 0.82 (0.07) | 0.86 (0.06) |
| 12 | Myo | 0.86 (0.08) | 0.81 (0.07) | 0.86 (0.07) |
| | LA | 0.93 (0.04) | 0.87 (0.03) | 0.93 (0.02) |
| | LV | 0.9 (0.04) | 0.84 (0.09) | 0.9 (0.05) |
| | RA | 0.91 (0.03) | 0.86 (0.03) | 0.92 (0.02) |
| | RV | 0.92 (0.01) | 0.89 (0.01) | 0.92 (0.01) |
| | AA | 0.96 (0.02) | 0.89 (0.04) | 0.95 (0.01) |
| | PA | 0.88 (0.05) | 0.79 (0.1) | 0.86 (0.06) |

Table 3. Euclidian distance of center of mass (COM) of the predicted brain structures with respect to the ground truth (mm). The network was trained on the full split of the dataset (i.e., 26 cases). Global binary masks served as the only input for the model.

|  | Brain Stem | Left eye | Right eye |
| --- | --- | --- | --- |
| **COM distance (mm)** | 4.7±1.5 | 2.7±1.3 | 2.0±0.7 |

Table 4. Euclidian distance of the center of mass (COM) of the predicted heart structures with respect to the ground truth (mm). The network was trained on the full split of the dataset (i.e., 12 cases). Global binary masks served as the only input for the model.

|  | Myo | LA | LV | RA | RV | AA | PA |
| --- | --- | --- | --- | --- | --- | --- | --- |
| **COM distance (mm)** | 1.9±0.3 | 3.8±1.4 | 1.5±0.7 | 3.5±1.6 | 2.4±1.5 | 2.7±0.8 | 1.5±0.7 |

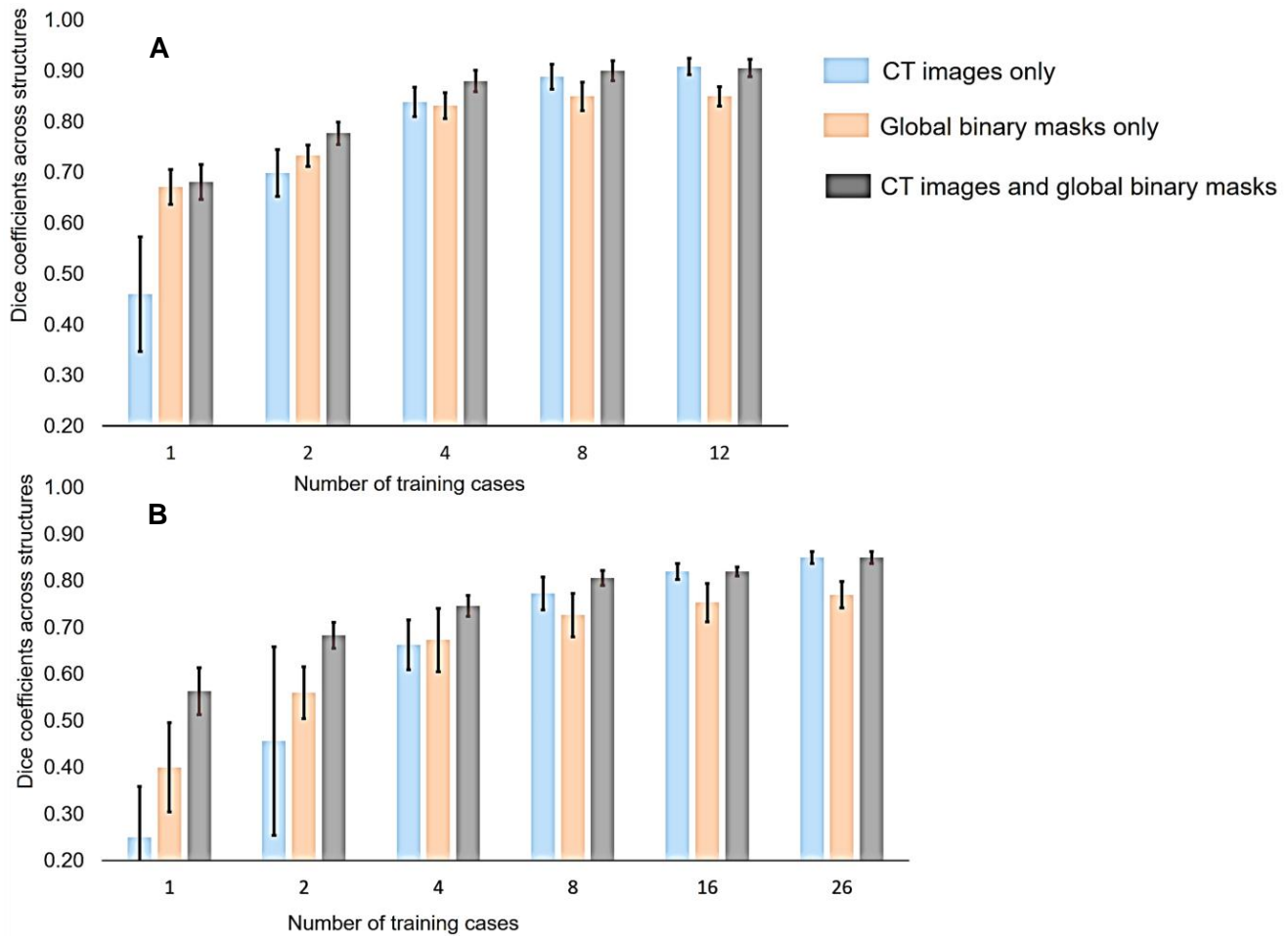

Figure 4. Average Dice coefficient across structures for the heart (A) and brain (B) datasets using different amounts of training data. The outcomes for the models trained on CT images only, global binary masks only, and the model with global binary masks as an additional channel are presented. Error bars represent standard deviation.

### 3.1.3. Quantitative results for reduced sets of training data

Utilizing global binary masks as the only input, provided DC of 0.4±0.19, 0.56±0.11, 0.67±0.14, 0.73±0.09, 0.75±0.08 across the brain structures when the model was trained on 1, 2, 4, 8, and 16 images respectively (Table 1). The model trained on CT images only appeared to outperform that of trained on global binary masks by 9.4%, 8.1%, and 6% for 26, 16, and 8 cases respectively (Figure 4A). However, as the amount of training data was reduced to 4, 2, and 1 cases, the proposed model showed more robust performance relative to the model trained on CT images; The model trained on global binary masks appeared to outperform that of trained on CT images by 1.5%, 22.6% and 60% using 4, 2, and 1 training cases. For 2 training cases, while the model trained on CT images totally failed to segment the right eye (DC = 0), global binary masks improved the accuracy to 0.64±0.15 (Table 1). The outcomes imply that in extremely small training data, encoding position and shape is superior to intensity information for segmentation.

For the heart dataset, the proposed model let to accuracies of 0.67±0.07, 0.73±0.04, 0.83±0.05, 0.85±0.06 using 1, 2, 4, and 8 training cases respectively (Table 2). The model trained on 12, 8, and 4 CT images showed improved performance relative to the model trained on global binary masks by 6.4%, 4.3%, and 0.9% respectively (Figure 4B). However, the difference between the two models' performance became less significant as lower amounts of training data was utilized. For instance, the models trained on 2 and 1 global binary masks outperformed that of trained on CT images by 4.9% and 46% respectively (Figure 4B); While the model trained on one CT image failed to segment LV (DC = 0), global binary mask improved DC for this structure to 0.69±0.05 (Table 2). Similar to the brain dataset, in the heart dataset, we observed that when access to training data is highly limited, encoding position and shape information is more effective than intensity information in making accurate inference. For example, the model exclusively trained on as few as 1 or 2 global binary masks was shown to be capable of providing rough segmentation of the structures and appeared to be more robust than that trained on CT images (Table 2).

### 3.2. Utilizing global binary masks as shape/position prior to assist segmentation

The effectiveness of incorporating global binary masks as an additional channel providing position/shape clues was studied for the two datasets. The outcomes were compared with the model trained on CT images for reduced sets of training data.

### 3.2.1. Qualitative evaluation

Figure 5, row A, represents examples of predicted masks by the proposed approach (third column) compared with the model trained on CT images only (second column) and the ground truth masks when the model trained on four training cases of the brain dataset. According to the outcomes, the segmentation masks produced by the model trained on CT images are inferior to those obtained by the proposed approach. For instance, using CT images as the only input, mistakenly produced negative predictions for image regions that belonged to the left and right eyes and the brain stem. This was more noticeable in segmentation of the brain stem where the model trained on CT images under-segmented this structure to a large extent. However, the proposed approach benefited from the information in the global binary masks which guided the network to learn more detailed information of target location and assisted the model for better segmentation leading to more accurate outcomes.

For the heart dataset, examples of test cases for the proposed model trained on four cases are presented in Figure 5, row B. Incorporating global binary masks led to improved segmentation relative to the model trained on CT images only. Inferior performance of the model trained on CT images only was more notable in segmentation of RV, RA, Myo, and LA; these substructures were highly under-segmented using this model. Using only CT images as the input, mistakenly produced negative predictions for image regions that belonged to the RV, RA, Myo, and LA. However, incorporating global binary masks provided shape/position information, guided the model for better performance, and let to relatively comparable outcomes to the ground truth masks despite training on only four cases.

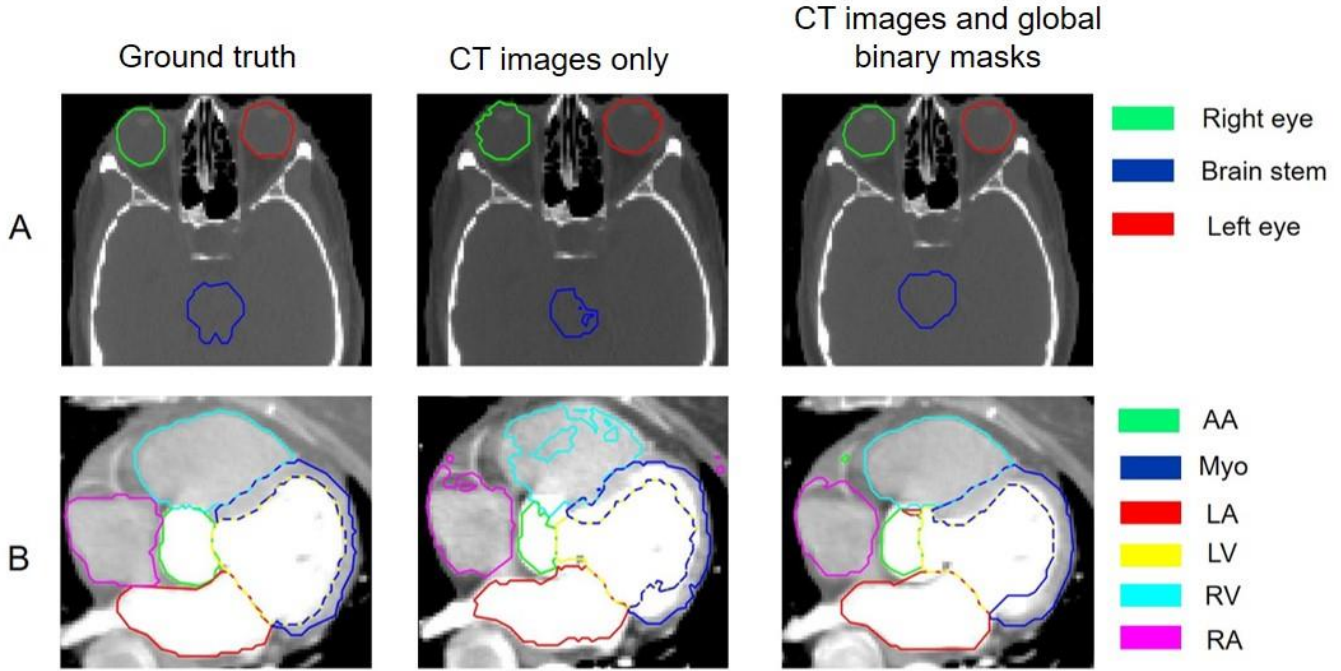

Figure 5. Qualitative comparison of the predicted contours by the model trained on CT images only (second column) and the model trained with CT images along with global binary masks (third column), i.e., the proposed second scenario, with respect to the ground truth contours (first column). Rows A and B are examples of test cases when the models were trained on 4 training cases for the brain and heart datasets respectively.

### 3.2.2. Quantitative results

Using full training set of the brain and heart datasets (i.e., 26 cases and 12 cases respectively), the performance of the model with global binary mask as an additional channel appeared to be identical to that of trained on CT images. Dice coefficients of 0.85±0.03 and 91±0.03 were obtained across structures for the brain and heart datasets, respectively (Table 1, Table 2). However, incorporating global binary masks significantly reduced the model's convergence time for both datasets (Figure 6). The training time for the model trained on 26 cases of the brain dataset decreased from 12 hours to 2 hours by addition of global binary masks (Table 5). Similar outcome was achieved for the heart dataset where the network convergence time dropped from 3.5 hours to 1 hour using the proposed approach (Table 6).

Table 5. Comparing the network training time (hours) for the model trained with CT images only versus the model with CT images along with global binary masks as an additional channel. Training time for different amounts of training data are presented for the brain dataset.

|  | Number of training cases | | | | | |
| --- | --- | --- | --- | --- | --- | --- |
|  | **1** | **2** | **4** | **8** | **16** | **26** |
| **CT images only** | 204h | 72h | 60h | 19h | 14h | 12h |
| **CT images and global binary masks** | 14h | 7.5h | 7h | 4h | 1.5h | 2h |

Table 6. Comparing the network training time (hour) for the model trained with CT images only versus the model trained CT images along with global binary masks as an additional channel. Training time for different amounts of training data are presented for the heart dataset.

|  | Number of training cases | | | | |
|---|---|---|---|---|---|
|  | 1 | 2 | 4 | 8 | 12 |
| **CT images only** | 15h | 10h | 6.5h | 5h | 3.5h |
| **CT images and global binary masks** | 12h | 9h | 4.5h | 2h | 1h |

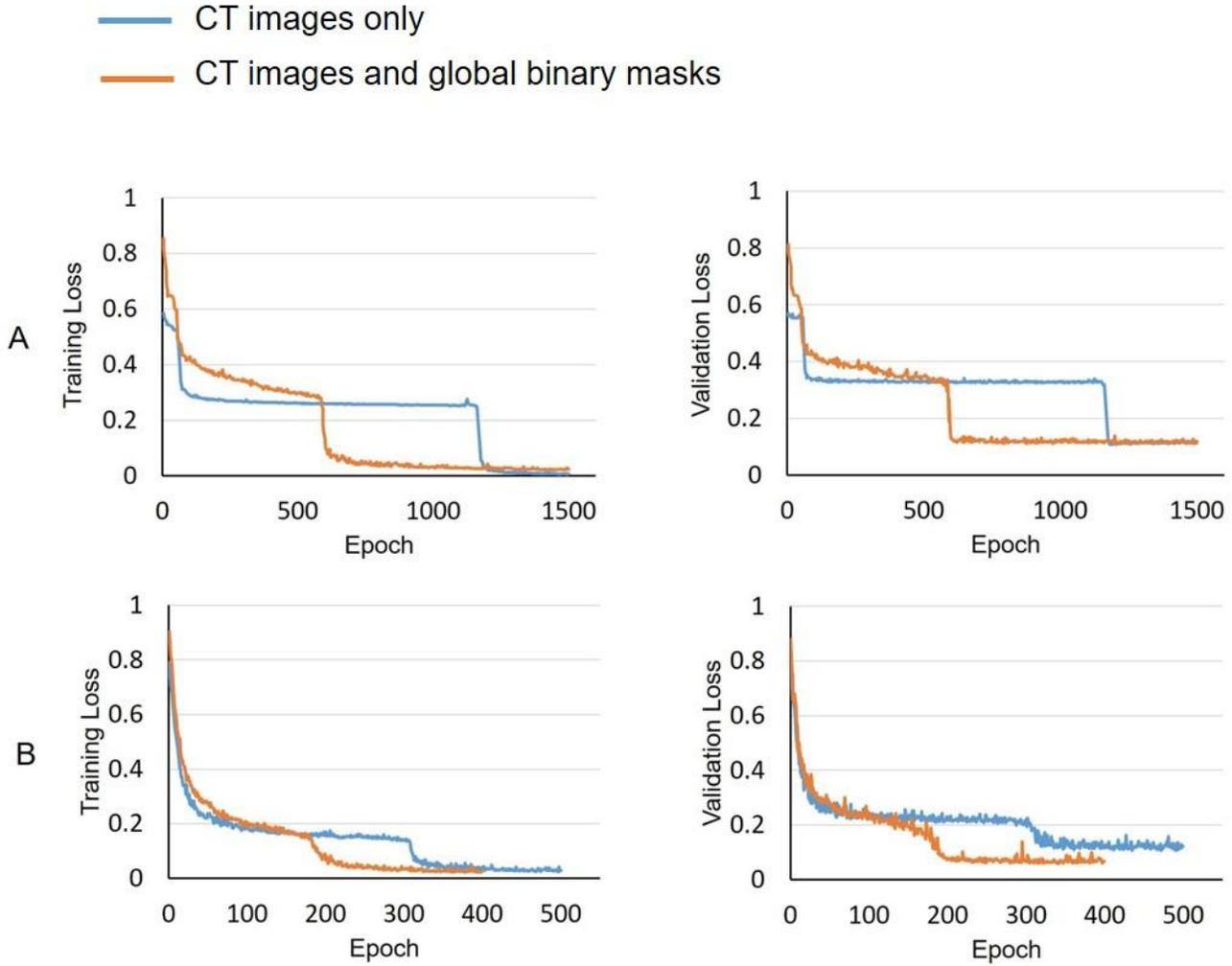

Figure 6. Comparing the convergence time of the model trained with CT images versus when the global binary masks were incorporated as an additional channel for the brain (A) and heart (B) datasets. The graphs present the training loss (left) and validation loss (right) using full training split. While providing equivalent performance, incorporating global binary masks led to remarkably less training time for both datasets.

### 3.2.3. Quantitative results for reduced sets of training data

For the brain dataset, average DC of 0.56±0.1, 0.68±0.06, 0.75±0.05, 0.81±0.03, and 0.82±0.02 were achieved for 1, 2, 4, 8, and 16 training cases respectively when global binary masks were utilized as the second channel (Table 1). For 8, 4, 2 and 1 training cases, incorporating global binary masks increased DC by 4.3%, 12.6%, 49.6%, and 125.3% respectively, relative to the model trained exclusively on CT images (Figure 4A). For 8 training cases, the proposed approach led to improved DC from 0.7±0.07 to 0.77±0.06 for the brain stem. Similar results were obtained using 4 and 2 training cases, where using only CT images as the input led to DC of 0.57±0.11 and 0.60±0.14 respectively, and the proposed approach improved the outcomes to 0.7±0.05, 0.68±0.07 for the brain stem. Furthermore, while the model trained on 2 CT images appeared to fail in segmenting the right eye (DC=0), utilizing global binary masks increased DC to 0.63±0.1 (Table 1). Using 16 training cases, the average accuracy across structures with global binary masks was the same as that of the CT images (0.82±0.02). However, incorporating global binary masks significantly reduced the model's convergence time (Table 5). The training time for the model trained on 16 cases decreased from 14 hours to 1.5 hours when global binary masks were incorporated. Overall, depending on the size of training data, utilizing the global binary masks led to achieving equivalent or better performance with remarkably less training time. For instance, for 1, 2, 4, and 8 training cases, global binary masks reduced the training time from 204h, 72h, 60h, and 19h to 14h, 7.5h, 7h, and 4h respectively along with providing superior segmentation (Table 5).

Similar to the brain dataset, for the heart dataset, the proposed approach outperformed that of trained on CT images by 4.9%, 11.2%, and 48.1% using 4, 2, and 1 training cases (Table 2, Figure 4B). For instance, while using CT images as the only input resulted in accuracies of 0.77±0.08 and 0.58±0.15, and 0.59±0.15 for segmentation of Myo using 4, 2, and 1 training cases, incorporating global binary masks increased DC to 0.84±0.08, 0.74±0.11, and 0.69±0.11 respectively. The model trained on CT images also failed to segment LV (DC=0) for one training case, however, the proposed model achieved DC of 0.66±0.09 for this structure (Table 2). The positive effects of global binary masks became less noticeable when larger amounts of training data were utilized (Table 2, Fig 4B). For instance, using 8 training cases the model trained on CT images and the proposed approach provided similar outcomes. However, even when incorporating global binary masks did not significantly improve the performance, it led to remarkably reduced model's convergence time (Table 6). For instance, the training time for the model trained on 8 cases decreased from 5 hours to 2 hours. Overall, depending on the size of training data, incorporating global binary masks led to achieving equivalent or better performance but with significantly less training time. For instance, using 1, 2, and 4 training cases, in addition to significantly higher performance, the training time was reduced from 15h, 10h, and 6.5h to 12h, 9h, and 4.5h respectively using the global binary masks (Table 6).

For both datasets, it was observed that the performance of the model trained on CT images deteriorated quickly as the model access to the training data reduced. It was then seen that this performance deterioration was mitigated when an additional channel embeds position information to the input images via global binary masks. The results support our hypothesis regarding global binary masks contributions to improvement in generalization over limited training data.

### 3.2.4. Feature maps visualization
### 3.2.4.1. Feature maps interpretation for the model with global binary masks
In this section, feature map visualization was provided to shed light on the types of features the model considers important for accurate segmentation of brain substructures. As an example, all 32 feature maps from the last double convolution layer of the 4$^{th}$ up-sampling block of the model trained on four CT images and global binary masks of the head were presented in Figure 7. Visualization of feature maps allows to identify which regions in the input CT image leads to high activations (Huff *et al* 2021, Yamashita *et al* 2018). High-activated regions in the feature maps are represented by higher intensity values which correspond to the areas that the model finds most relevant for the segmentation task (Lecun *et al* 2015, Zhang and Zhu 2018). According to Figure 7, the following types of features appeared to be valuable for the prediction of brain substructures (i.e., the right and left eyes and the brain stem):

**Substructure-specific features:** The high intensities observed in the feature maps corresponding to the left and right eye and the brain stem regions (e.g., 5, 10, 11, 19, 25) suggest that the model is effectively capable of identifying and localizing the substructure-specific patterns. These patterns may include characteristic shapes, curvatures, intensities, and spatial arrangements associated with the substructures allowing the model to segment them accurately (Kamnitsas *et al* 2017, Zhang and Zhu 2018)

**Multi-region features:** Joint activation of different substructures in a feature map (e.g., 17, 26, 29) suggests that the U-Net model has acquired valuable shared representations. Joint activation of the substructures in a feature map indicates that the model has learned to capture common knowledge, shared features, and relationships between these anatomical structures, which can be valuable for accurate and robust segmentation (Zhou *et al* 2016). The varying intensities in the shared feature maps reflect the model's segmentation confidence for each region (Zeiler and Fergus 2014). Higher intensities at one region compared to another could indicate stronger confidence in the segmentation of that particular structure (e.g., the left eye versus brain stem in feature map 17).

**Contextual features:** The activations observed on the concepts that network was not trained to learn such as the whole head region (e.g., 3, 4, 28) and the areas around the nasal cavity (e.g., 2, 12, 22), is indicative of the model's ability to recognize global patterns and contextual cues. These features could help the model understand the overall layout of the head in relation to the segmented regions (Kamnitsas *et al* 2017).

**Background activations:** The feature maps with activations on the background regions (e.g., 2, 12, 22, 24), can demonstrate the model's capability to distinguish between foreground and background elements in the input image. Such features play an essential role in excluding irrelevant regions during segmentation (Van Molle *et al* 2018).

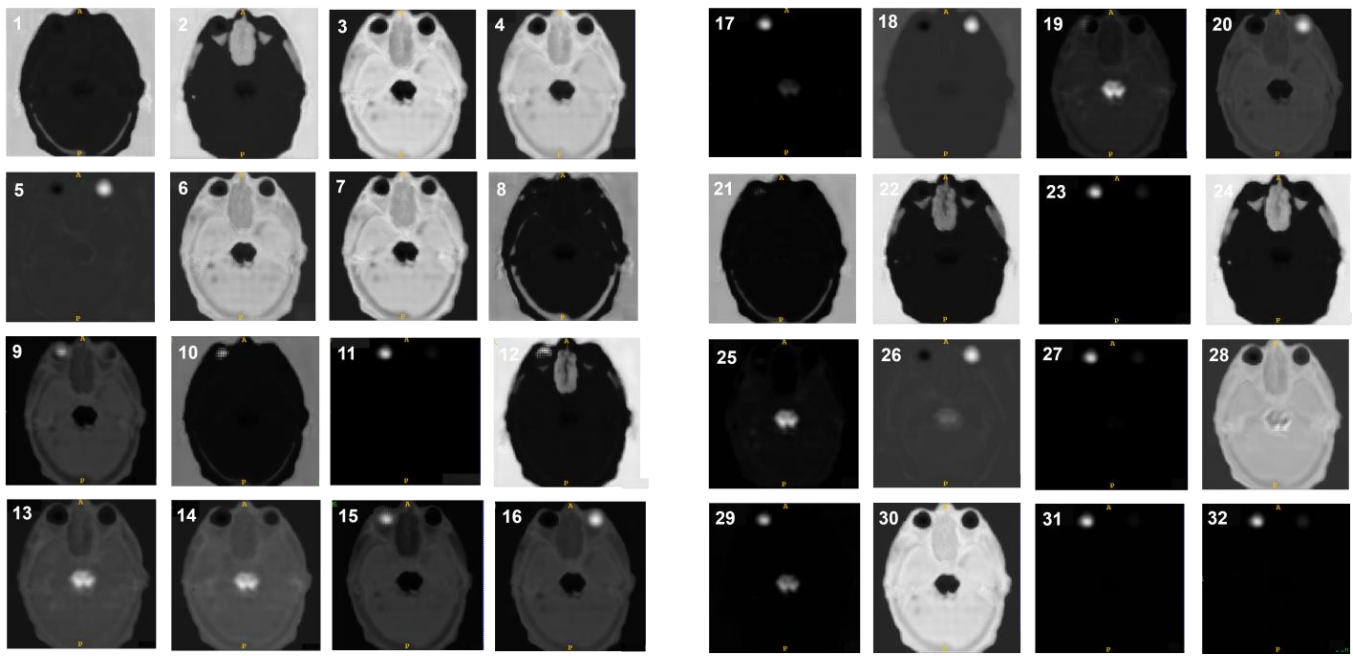

Figure 7. The 32 feature maps extracted from the last double convolution layer of the 4[th] up-sampling block of the network. The network trained on four CT image plus global binary masks to segment brain substructures including the right and left eye and the brain stem.

### 3.2.4.2. Feature maps comparison across models

Here, we attempt to extract visualizations of the learnt feature maps for the networks trained on only brain CT images and compare that with when global binary masks were incorporated as an additional channel (Figure 8). Feature maps were obtained at the last double convolution layer of the 4$^{th}$ up-sampling block of the network. Rows A, B, and C, represent example feature maps learnt by the models trained on 4 CT images only, 4 CT images and global binary masks, and 8 CT images respectively. Columns 1-8, represent example feature maps learnt by the individual filters of the networks. Comparing rows A and B implies that incorporating global binary masks leads to extracting features maps that represent brain substructures (i.e., brain stem, left eye, right eye) more accurately relative to the network with one channel input. For instance, feature maps in row A, columns 1 and 8, had relatively low activation on the brain stem, while the corresponding feature maps in row B, showed significantly higher activation on the brain stem. Comparing feature maps of rows A and B revealed significantly finer segmentation of the whole head and the structures when global binary masks were incorporated. Higher attention of the network on the whole head was clearly observable in columns 2, 3, 6, 7, and 8. In column 5, row B, a clear segmentation of the whole head was observed, while in row A, a large area of the whole head showed similar intensity levels to the background. This could be indicative of the role of global binary mask in suppressing the influence of non-informative background leading to better segmentation outcomes. Similarly, left and right eyes were segmented with significantly higher accuracy in features maps of columns 2, 3, 6, 7, and 8. Furthermore, feature maps learnt by the network trained on 4 CT images and the global binary masks appeared to be similar to that of trained on 8 CT images only (row B versus C). This supports our second hypothesis on the benefits of incorporating global binary masks to compensate for scarcity of the training data.

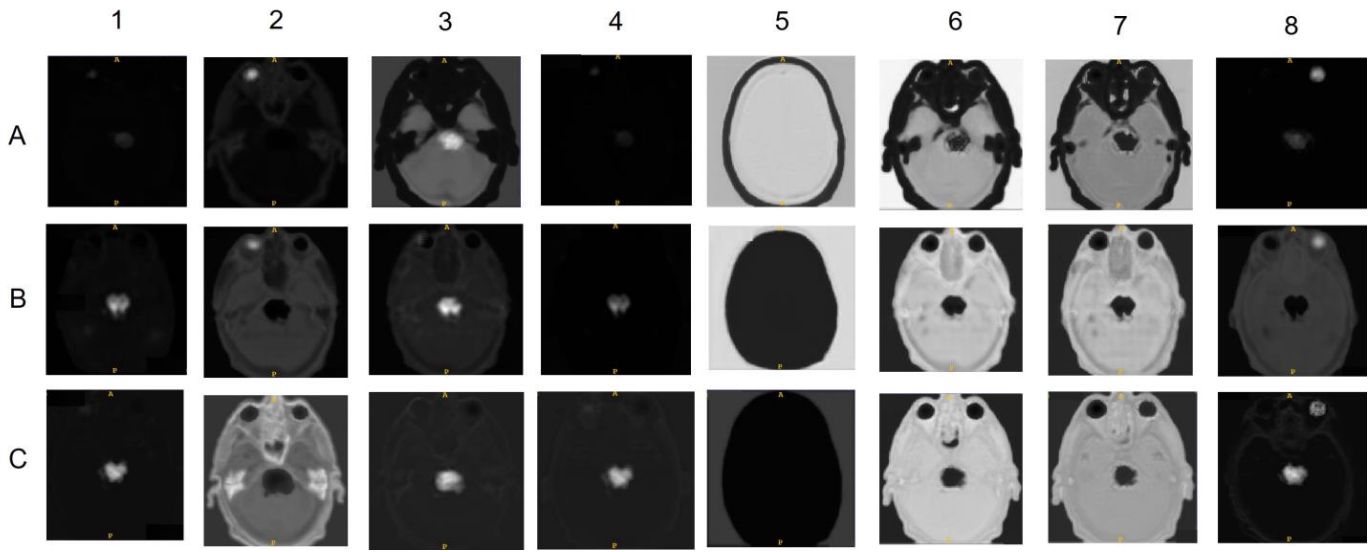

Figure 8. Feature maps visualizations for the networks trained on only 4 CT images (A), 4 CT images and global binary masks (B), and 8 CT images only. Columns represent example feature maps learnt by the individual filters at the last double convolution layer of the 4$^{th}$ up-sampling block of the networks.

### 4. Discussion

Despite the promising performance of deep learning approaches for medical image segmentation, their deployment into clinical practice can result in severe performance degradation due to domain shift across imaging protocols, acquisition parameters, and image source (De Fauw *et al* 2018, Zhang *et al* 2020, Abràmoff *et al* 2018). The Image appearance and quality can be variable in such circumstances. Hence, poor generalizability of deep learning models is one of the major issues hindering their deployment into realistic clinical environment (Yasaka and Abe 2018). Common domain adaptation and transfer learning based methods proposed to address the generalization issue, require to access the target domain data for re-training the model (Karimi *et al* 2021) , which is not feasible for every new application. In contrast, for human it is quite easy to identify the same organ in different modalities as human eye predominantly incorporates shape and position information to identify the objects. For example, by showing a CT scan of the brain and pointing out the eyes, even a non-expert will be able to identify the same organs in MRI images. Our proposed approach was inspired by the way the human eye localizes objects; we proposed to utilize position and shape information of the organs to perform segmentation tasks.

Our proposed approach in the first scenario provided average DC of $0.77\pm0.06$ and $0.85\pm0.04$ when the models were exclusively trained on 12 and 26 global binary masks of the head and heart respectively. The results indicated that a surprising degree of position and shape information is encoded when we remove the model access to intensity information by replacing global binary masks for CT images. The approach has the potential to overcome the generalizability issues as it is totally independent of image intensity and modality. Such setting has implications for a wide range of applications in unseen domains such as different scanners, acquisition sessions and image quality as only global binary masks are required for inference. Hence, the approach could eliminate the requirement of including target domain data in the training step. However, it should be considered that atypical organs' anatomy could present great challenges for the model trained on global binary masks only, as this model has been trained to learn the typical spatial, location, and size of structures. For instance, in the case of a diseased organs, organs with transplants, or surgical clips the model might fail in generating accurate contours and might provide inferior performance. While this framework may not be ideal for detailed segmentation of the structures, it has the potential to be employed for organ localization and rough segmentation. Relatively small variations of the center of mass of the predicted masks relative to the ground truth for different structures (Table 3 and 4) confirm the implications of the proposed method for structure/organ localization in medical images. Organ localization and rough segmentation have been reported to be of crucial importance in many image segmentation tasks (Xu *et al* 2019, De Vos *et al* 2017). For instance, by detecting the target organs and identifying their corresponding bounding boxes, non-relevant information were shown to be discarded during segmentation task to focus on regions of interest and process these regions locally. Target localization approaches have also been reported to improve performance of the segmentation models compared to the current state-of-art, while requiring significantly less computational resources and reduced the inference time (Shirokikh *et al* 2021).

To boost the performance of segmentation model trained with insufficient data, our proposed approach for incorporating global binary masks as an additional channel was demonstrated to be effective. When the model trained on only CT images failed to accurately segment the structures due to lack of sufficient training data, utilizing global binary mask led to encoding more precise features and thus improved performance Table 1, Table 2). We combined feature maps from the CT images and those from the global binary mask to provide position clues for the model. This leads to more precise target structure localization in small datasets and was shown to improve robustness to insufficient training data relative to the model trained on CT images only. For instance, in the heart dataset trained on one case only, without global binary masks, the model totally failed to segment LV (DC=0) while incorporating global binary masks increased the DC to $0.66\pm0.09$ (Table 2). Similarly, for 2, 4 and 8 training cases, incorporating global binary masks improved the average DC across structures from $0.7\pm0.09$, $0.84\pm0.06$, and $0.89\pm0.05$ to $0.78\pm0.04$, $0.88\pm0.04$, and $0.90\pm0.04$ respectively. Similar results were obtained for brain dataset; Utilizing global binary masks increased the DC from 0 to $0.64\pm0.15$ for segmentation of right eye using two training cases (Table 2). For 4 and 8 training cases, the average DC across structures improved from $0.66\pm0.11$ and $0.77\pm0.07$ to $0.75\pm0.05$ and $0.81\pm0.03$ respectively. For the heart dataset, it was also observed that incorporating global binary masks for the model trained on 4 cases provided similar accuracy to that of trained on 8 cases without global binary masks. For the brain dataset, training on 4 cases with global binary masks provided similar accuracy to that of trained on 8 cases without global binary masks. The same trend was reported for 8 and 16 images for this dataset. All these results indicate that integrating domain knowledge into the deep learning model in the form of global binary masks of the heart and head can compensate for the scarcity of the training data to a large extent. This is of great significance in medical images as annotating medical images is a time-consuming, labor-intensive, and expensive process. For instance, manual delineation of whole heart requires almost 8 hours for a single subject (Zhuang and Shen 2016), and the lack of large and annotated datasets has been identified as the primary limitation of the application of deep learning for segmentation tasks (Esteva *et al* 2019).When the size of training data is insufficient, the U-Net model trained with CT images only, cannot achieve an ideal performance which could create room for improvement by incorporating global binary masks as an additional input. Expanding the size of training data increases the intensity information the model has access to, which reduces model reliance on the position information provided by the global binary masks. This could explain the increasing similarity between the performance of the model trained exclusively on CT images and the proposed approach in the second scenario when training data expands. Nonetheless, incorporating global binary masks appeared to remarkably reduce the model's convergence time even for the larger amounts of training data for which utilizing global binary masks did not improve the accuracy (Figure 6, Table 5, Table 6). Although accelerating the training process of deep neural networks has been often under-appreciated in the medical image segmentation tasks, we argue that a good deep learning model should provide good performance with less training time (Wang *et al* 2019).

In this study, two different approaches leveraging global binary masks were comprehensively evaluated for structure segmentation and localization in medical images. Depending on the application, the type, and the amount of training data available, the most appropriate approach should be utilized. For instance, the model with global binary masks as an additional input can be desirable when the goal is to maximize the accuracy despite highly limited access to labeled training data. For multimodality training data or where there is a large domain shift between the training and test sets, the setting with the global binary masks as the only input could be advantageous. This is due to the reason that this approach is designed to learn domain-invariant representations by minimizing the domain discrepancy across multiple source domains via training on purely binary global binary masks. Furthermore, each proposed scenario could be used an initial step to provide rough segmentation where a refine model is then incorporated for accurate segmentation (Chen *et al* 2019). For instance, using the proposed approach, the whole image can be roughly processed to identify areas of interest, then locally segment each small region independently.

The proposed structures utilize the global binary masks to guide the fine structure segmentation, thus poorly generated binary mask could lead to poor segmentation results. However, generating accurate binary masks either by thresholding the CT image (e.g., the head dataset) or using a CNN/U-Net network (e.g., the heart dataset) is a straightforward process in the settings of this work. Both scenarios are very unlikely to produce inaccurate masks. This is due to the reason that the foreground in such settings has a highly distinctive boundary and there is high contrast between the foreground and the background. Besides, the relatively consistent size and shape of the foreground across each dataset makes the segmentation task straightforward for the network.

## 5. Conclusion

In this study, we investigated the advantages of leveraging global binary masks for structure localization and segmentation in medical images by proposing two scenarios. In the first scenario, using exclusively global binary masks for model training appeared to be robust and thus is generalizable across image modalities. In the second scenario, we demonstrated that leveraging global binary masks is an effective approach to compensate for scarcity of training data to a large extend, by significantly improving segmentation accuracy where access to labeled data is highly limited. The approach also highly reduced the training time for varying amounts of training data.

**Acknowledgement**

This work was partially supported by the National Institutes of Health under Grant No. R01-CA235723, 9R44CA254844-02 and 75N91021C00031.